\documentclass[superscriptaddress,groupedaddress,nofootnoteinbib,11pt]{article}  

\usepackage[usenames,dvipsnames]{color}
\usepackage{graphicx}  
\usepackage{dcolumn}   
\usepackage{bm}        
\usepackage{jcappub}

\def\be{\begin{equation}}
\def\ee{\end{equation}}
\def\ba{\begin{eqnarray}}
\def\ea{\end{eqnarray}}

\def\L*{{\cal L}_*}
\def\L{\mathcal{L}}
\def\({\left(}
\def\){\right)}

\def\<{\langle}
\def\>{\rangle}

\def\cs2{c_{s}^{2}}

 \def\ga{\gamma}

 \def\la{\lambda}

\def\gmn{g_{\mu\nu}}

\def\ggmn{G_{\mu\nu}}

\def\sigmamnd{\nabla_{\mu}\sigma\nabla_{\nu}\sigma}

\def\dij{\mathcal{D}_{ij}}
\def\dsg{\delta\sigma}

\def\la{~\mbox{\raisebox{-.6ex}{$\stackrel{<}{\sim}$}}~}
\def\ga{~\mbox{\raisebox{-.6ex}{$\stackrel{>}{\sim}$}}~}

\begin{document}

\title{Gravitational Waves and Scalar Perturbations  from Spectator Fields}

\author{\hspace{-6pt}Matteo Biagetti$^a$,}
\author{Emanuela Dimastrogiovanni$^{b,c}$,}
\author{Matteo Fasiello$^{d,e}$,}
\author{Marco Peloso$^b$ .}
\affiliation{$^a$D\'epartement de Physique Th\'eorique and Centre for
Astroparticle Physics (CAP),\\ \textcolor{white}{} Universit\'e de Gen\`eve, 24 quai E. Ansermet, CH-1211 Gen\`eve, Suisse}
\affiliation{$^b$ School of Physics and Astronomy, University of Minnesota, Minneapolis, MN 55455, USA}
\affiliation{$^c$Department of Physics and School of Earth and Space Exploration, Arizona State University, Tempe, AZ 85827, USA}
\affiliation{$^d$ CERCA \& Department of Physics, Case Western Reserve University,  Cleveland, USA}
\affiliation{$^e$ Stanford Institute for Theoretical Physics,
Stanford University, Stanford, CA 94306, USA}
\date{\today}
\leftline{UMN-TH-3408/14}

\abstract{
The most conventional mechanism for gravitational waves (gw) production during inflation is the  amplification of vacuum metric fluctuations.  In this case the gw production can be uniquely related to the inflationary expansion rate $H$. For example, a gw detection close to the present experimental limit (tensor-to-scalar ratio  $r \sim 0.1$) would indicate an inflationary expansion rate close to $10^{14} \, {\rm GeV}$. This conclusion,  however,  would be invalid if the observed gw originated from a different source.  We construct and study one of the possible covariant formulations of the mechanism  suggested in  \cite{Biagetti:2013kwa}, where a spectator field $\sigma$ with a sound speed $c_{s} \ll 1$ acts as a source for gw during inflation. In our formulation $\sigma$ is described by a so-called $P(X)$ Lagrangian and a non-minimal coupling to gravity. This field interacts only gravitationally with the inflaton, which has a standard action. We compute the amount of scalar and tensor density fluctuations produced by $\sigma$ and find that, in our realization,  $r$ is not enhanced with respect to the standard result but it is strongly sensitive to $c_s$, thus breaking the direct $r \leftrightarrow H$ connection.   \\ }
\maketitle


\section{Introduction}

Inflation \cite{Guth:1980zm,Lyth:1998xn,Riotto:2002yw,Kinney:2003xf} is a central paradigm for the physics of the very early universe; already in its simplest realization, with a single slowly-rolling field and a canonical Lagrangian, it delivers predictions which are in excellent agreement with observations \cite{Ade:2013uln}. During inflation gravitational waves (henceforth ``gw") are produced from the vacuum fluctuations of the tensor modes of the metric, and they become classical as they are stretched outside the Hubble radius and reach the constant power 
\begin{equation}\label{one}
\mathcal{P}_{h}=\frac{2\,H^{2}}{\pi^{2}M_{P}^{2}},\quad\quad\quad r \sim16\,\epsilon. 
\end{equation}
In the first relation, $H$ denotes the Hubble expansion rate, and $M_p$ the reduced Planck mass. The second expression relates the tensor-to-scalar ratio, $r \equiv \frac{\mathcal{P}_{h}}{\mathcal{P}_{s}}$ , to the slow roll parameter $\epsilon \equiv \frac{M_p^2}{2} \, \left( \frac{V'}{V } \right)^2$. From here, one also obtains the well know Lyth relation \cite{Lyth:1996im}
\begin{equation}\label{two}
\frac{\Delta\phi}{M_{P}}= \sqrt{2\,r}\Delta N,
\end{equation}
in the case of single field inflation. This expression gives the field range $\Delta \phi$ spanned by the inflaton during $\Delta N$ e-folds of inflation. 

Primordial gw induce B-mode polarization fluctuations in the Cosmic Microwave Background (CMB) \cite{Seljak:1996gy,Kamionkowski:1996zd}. CMB B-mode were recently detected by BICEP2 \cite{Ade:2014xna,Ade:2014gua} 
and POLARBEAR \cite{Ade:2014afa}. In particularly, the BICEP2 signal is in a multipole range that can be affected by primordial gw, although its astrophysical origin seems the more likely interpretation at this moment \cite{Flauger:2014qra,Cortes:2014nqa,Adam:2014bub}.  As the observed CMB modes were produced during  $\Delta N \sim 5$ e-folds of inflation  \cite{Lyth:1996im}, eq. (\ref{two}) indicates that an observation of gw in the near future (only possible if $r \ga 0.01$ \cite{Baumann:2008aq}) would rule out ``small field'' models of single field inflation, namely models in which $\phi < M_p$ during inflation \cite{Antusch:2014cpa}. 

Clearly, these conclusions rely on the  standard vacuum amplification mechanisms for tensor and scalar perturbations during inflation (for a discussion see \cite{Kehagias:2014wza}), and on the fact that the gw are the main cosmological source of the CMB B-modes. Alternative cosmological sources have been studied in the literature, including  primordial magnetic fields \cite{Bonvin:2014xia}, phase transitions \cite{Krauss:1991qu,Kamionkowski:1993fg,JonesSmith:2007ne,Krauss:2010df,Dent:2014rga,Durrer:2014raa}, and topological defects \cite{Lizarraga:2014eaa,Moss:2014cra}.

Although quite resilient (see \cite{Creminelli:2014wna}), even in the context of inflation the conditions (\ref{one}-\ref{two}) may be invalid  in presence of extra sources for scalar and / or tensor perturbations.
For instance,  models have been proposed in which gw are produced from particles (and strings) that are generated during the inflationary era through either sudden particle production or through copious generation of vector quanta
 \cite{Green:2009ds,Anber:2009ua,Barnaby:2010vf,Sorbo:2011rz,Barnaby:2011vw,Cook:2011hg,Senatore:2011sp,Barnaby:2012xt,Carney:2012pk,Cook:2013xea,Shiraishi:2013kxa,Mukohyama:2014gba,Ferreira:2014zia,Ozsoy:2014sba}
  \footnote{Production of gw in inflationary models where the initial states for scalar and tensor perturbations are non Bunch-Davies has also been considered (see e.g. \cite{Hui:2001ce,Collins:2014yua,Aravind:2014axa}).}. The major challenges for these models normally resides in being able to ensure that an observably large production of gw has no dangerous implications for the scalar sector \cite{Barnaby:2012xt}. Whatever the gw source is, it couples to gw with gravitational strength. It will also couple with comparable or higher strength to the  field(s) responsible for the observed scalar perturbations. Therefore a strong gw source does not necessarily implies a growth of the observable ratio~\footnote{In adding the power spectra we are using the fact that the vacuum and the sourced signal are statistically uncorrelated.} 
\begin{equation}
r = \frac{P_{h,{\rm vacuum}} + P_{h,{\rm sourced}}}{P_{\zeta,{\rm vacuum}} + P_{\zeta,{\rm sourced}}} 
\label{r}
\end{equation}
once the increase of the scalar power is also accounted for.~\footnote{This is for example the case in models of warm inflation \cite{Berera:1995ie}, as shown in \cite{BasteroGil:2009ec}.}  Moreover, the sourced modes are in general expected to be non-Gaussian, and therefore the observed Gaussianity of the CMB anisotropies \cite{Ade:2013uln}
typically results into strong limits on any given mechanism of particle production. Rather ad-hoc constructions typically need to be advocated if one wants to both respect these limits and generate a visible gw signal \cite{Barnaby:2012xt}.   
 
 Another possibility has been explored in \cite{Biagetti:2013kwa}, where gw are sourced by the fluctuations of a spectator scalar field during inflation (see \cite{Ananda:2006af} for more models where tensor modes are generated by scalar fluctuations). In general, tensor modes that are generated at second order in perturbation theory are expected to be suppressed with respect to  the vacuum perturbations. However, production at non-linear order may be enhanced in models where the sourcing scalar field has a small sound speed. This is precisely the scenario presented in \cite{Biagetti:2013kwa}. In this model, inflation is driven by a single scalar field, in the presence of a spectator field. The latter provides negligible contribution to the total energy density of the universe during inflation and also negligible direct contribution to the power spectrum of curvature fluctuations. The spectator field has a sound speed $c_{s}<1$, and the action at second order for the field fluctuations $\delta\sigma$ is
\begin{equation}\label{chico}
\mathcal{S}_{\delta\sigma}^{(2)}=\int d^{3}x \, d\tau \, a^{4}\left[\frac{1}{2 a^{2}}\left(\delta\sigma^{'2}-c_{s}^{2}\left(\nabla\delta\sigma\right)^{2}\right)-V[\delta\sigma]\right].
\end{equation}
The authors of  \cite{Biagetti:2013kwa} computed the contribution to the power spectrum of gw arising at second order from the convolution of $\sigma$ fluctuations. In this setup the tensor power spectrum is easily enhanced; the model, unlike the standard case, also allows for a blue spectrum (see \cite{Cannone:2014uqa} for another very recent proposal). 

To ascertain whether this gw production may result in a greater observable $r$, the ratio (\ref{r}) should be computed in explicit realizations of this model. The first step in this computation is to find a possible covariant formulation for the model in \cite{Biagetti:2013kwa}. To describe the dynamics of the spectator field $\sigma$, we employ a Lagrangian of the type  $P(X,\sigma)$ , where $X \equiv - \left( \partial \sigma \right)^2$; a sound speed smaller than unity is easily implemented in these models.
Indeed, a large class of top-down realizations of inflationary mechanisms allows for fields whose speeds of sound are smaller than unity, often times the inflaton field itself \cite{Alishahiha:2004eh} (see also \cite{ArmendarizPicon:1999rj} for more phenomenologically oriented models). From a high energy theory standpoint, the presence of several light fields at energies of order $E_{infl}$ or higher is also plausible and we note here that, depending on the coupling (see e.g. \cite{Tolley:2009fg,Achucarro:2010jv}), integrating out of these fields can lead to a modified speed of propagation for the remaining field(s).

For the theory at hand, we also assume a non-minimal derivative coupling of $\sigma$ to gravity~\footnote{Later in the text, we will elaborate on the reasons behind this choice.}. The spectator sector is minimally coupled to an inflaton field $\phi$ that has a standard Lagrangian, and hence a unitary sound speed. We compute the contributions to scalar and tensor power spectra due to second order contributions sourced by $\sigma$.

We find that in our setup $r$ is not enhanced compared to the standard generation mechanism but that it is highly dependent on the value of the speed of sound $c_s$ for a large region of the parameters space. The direct $r \leftrightarrow H$ relation is therefore broken. 

This paper is organized as follows: in Sec.~2 we discuss our model, study the background evolution and the perturbations for the spectator field; in Sec.~3 and 4 we present the main steps of the calculation of the second order contributions from $\delta\sigma$ to the tensor and scalar power spectra; in Sec.~5 we summarize our results and we comment on our findings; in Sec.~6 we offer our conclusions. More details about the full linear perturbation analysis and about the computation of the non-linear equation for the scalar fluctuations are provided respectively in Appendices~\ref{A} and \ref{B}. Details about tadpole diagrams are presented in Appendix~\ref{C}.

\section{A covariant description of a field with $c_{s}<1$}
\label{cov}

In our set-up inflation is driven by a scalar field $\phi$ minimally coupled to gravity and to a spectator field $\sigma$. The latter has a sound speed $c_{s}<1$. The Lagrangian for $\sigma$ is also characterized by a non-minimal derivative coupling to gravity \footnote{Inflationary Lagrangians with non-minimal derivative couplings to gravity for the scalar fields were proposed by several authors \cite{Horndeski:1974wa}. In our model, it is only the spectator field that enjoys a non-minimal coupling to gravity.}. The total action reads
\begin{equation}\label{totala}
S = \int d^4 x \sqrt{-g} \,\,\, \left\{\frac{M_{P}^{2}}{2}R-\frac{1}{2}g^{\mu\nu}\partial_{\mu}\phi\partial_{\nu}\phi-V(\phi)+\mathcal{L}_{\sigma} \right\},
\end{equation}
(the signature of the metric is $-+++$) where 
\begin{equation}\label{totalas}
\mathcal{L}_{\sigma} \equiv \Lambda^{4}\left(-\frac{g^{\mu\nu}\partial_{\mu}\sigma\partial_{\nu}\sigma}{M^{4}}\right)^{n}-\frac{\Lambda^{2}}{M^{4}}G^{\mu\nu}\partial_{\mu}\sigma\partial_{\nu}\sigma .
\end{equation}
$G_{\mu\nu}$ is the Einstein tensor, $\Lambda$ and $M$ are constant energy scales and $n>1$. The field $\sigma$ has no potential. As shown explicitly later in the linear perturbation analysis, one can verify that the minus sign accompanying the non-minimal coupling term ensures the absence of instabilities. \\
\indent  Assuming a negligible contribution from the spectator field to the total energy density, the spectator field does not affect the background evolution during inflation, nor it contributes directly to curvature perturbations. Our focus will lie uniquely on the spectator field fluctuations acting as a non-linear source for the scalar and tensor power spectra.   \\
\indent Einstein's equations read
\begin{eqnarray}
&&3H^{2}M_{P}^{2}=\frac{\dot{\phi}^{2}}{2}+V(\phi)+\rho_{\sigma},\nonumber\\
&&\dot{H}\left[1-\frac{n}{n-1}\frac{\rho_{\sigma}}{3H^{2}M_{P}^{2}}\right]=-\frac{\dot{\phi}^{2}}{2M_{P}^{2}}\, ,
\end{eqnarray}
where $\rho_{\sigma}$ is the contribution to the total energy density from the spectator field. The effect of $\sigma$ on the background and inflaton evolution is negligible provided that $\rho_{\sigma}\ll 3H^{2}M_{P}^{2}$.  \\
The background dynamics of the spectator field is governed by 
\begin{equation}
\frac{a^{3}\dot{\sigma}}{M^{2}}\left[n\Lambda^{2}\left(\frac{\dot{\sigma}^{2}}{M^{4}}\right)^{n-1}-3H^{2}\right]=const.
\end{equation}
One can use the ansatz
\begin{equation}\label{anz}
\dot{\sigma}=M^{2}\left[\sqrt{\frac{3}{n}}\frac{H}{\Lambda}\right]^{\frac{1}{n-1}},
\end{equation}
from which the energy density of the spectator field becomes
\begin{equation}\label{cen}
\rho_{\sigma}=-(n-1)\left(\frac{3}{n}\right)^{\frac{n}{n-1}}\left(\frac{\Lambda^{2}}{H^{2}}\right)^{\frac{n-2}{n-1}}H^{4}.
\end{equation}

We note that the energy density of the spectator field is nearly constant; one can study also the theory (\ref{totala}) in the absence of the non minimal coupling to gravity (namely, without the second term in (\ref{totalas})). Doing so, we obtained a background energy density $\rho_\sigma \propto a^{-3 \left( 1 +\frac{1}{2n-1} \right)}$, and a sound speed $c_s^2 = \frac{1}{2 n - 1}$. A small sound speed can be obtained for $n\gg1$, but this then implies  $\rho_\sigma \propto a^{-3}$ during inflation. On the other hand, the energy density in the inflaton field is (nearly) constant. This would force us to place more restrictions on the model, such as assuming that the observable range of e-foldings corresponds to a period of time during which the spectator field energy density was already subdominant, but not yet decayed away. This is the reason that prompted us to include the second term in (\ref{totalas}).  A nearly constant solution for $\sigma$ also has a strong calculational advantage: one expects that a second order Lagrangian with a form similar to Eq.~(\ref{chico}) will follow if (\ref{anz}) applies, which leads to standard Hankel mode functions for the spectator field fluctuations. These can be easily integrated when computing the sourced power spectra.\\

The full linear perturbation analysis for the model, including the metric fluctuations, can be found in Appendix~\ref{A}. The inflaton and the spectator field inevitably couple through gravity already at linear level; however, as a result of our working assumptions (i.e. a negligible contribution to the total energy density from the spectator field and the usual slow-roll approximation), the mixing terms can be safely neglected. As a result, the second order Lagrangian for $\sigma$ is
\begin{equation}
\mathcal{S}_{\sigma}^{(2)}=\frac{6(n-1)\Lambda^{2}}{M^{4}}\int d\tau d^{3}k a^{2} H^{2}\left[|\delta\sigma_{k}^{'}|^{2}-\frac{\epsilon k^{2}}{3(n-1)}|\delta\sigma_{k}|^{2}\right]\,,\quad\quad \epsilon\equiv-\frac{\dot{H}}{H^{2}}.
\end{equation}
Normalizing with a Bunch-Davies vacuum, the leading-order solution for the spectator field fluctuations reads
\begin{equation}\label{sfmf}
\delta\sigma_{k}(\tau)=\frac{B}{k^{3/2}}(1+i\,c_{s}k\tau)e^{-i\,c_{s}k\tau},\quad\quad\quad B\equiv\frac{M^{2}}{\Lambda}\frac{1}{2\sqrt{2}\sqrt{\epsilon\,c_{s}}}.
\end{equation}
As is the case for the inflaton fluctuations, the leading-order tensor quadratic Lagrangian has the form that one would obtain in the absence of any spectator field (see Appendix~\ref{A}). \\

\indent It is well known that scalar, tensor and vector modes mix beyond linear order in perturbation theory; as a result, tensor modes can for example be sourced at second order by scalar fluctuations. This is precisely the production mechanism that will be explored within this model, the spectator field acting as a source for tensor and scalar fluctuations. \\
The inflaton and tensor modes, after canonical normalization, obey the following equation
\begin{equation}\label{eeom1}
\left[\partial_{\tau}^{2}+\left(k^{2}-\frac{a''}{a}\right)\right]Q_{X}(\tau,\vec{k}) = J_{X}(\tau,\vec{k}),\quad\quad X=\{\phi,\lambda=+,\lambda=- \}
\end{equation}
where $\lambda$ stands for the tensor modes polarization. In our specific model, a major contribution to the source term $J(X)$ is given by the convolution of two $\delta\sigma_{k}$ modes 
\begin{equation}
J_{X}(\tau,\vec{k})\equiv\int\frac{d^{3}p}{(2\pi)^{3/2}}\hat{O}_{X}(\tau,\vec{k},\vec{p})\delta\sigma_{\vec{p}}(\tau)\delta\sigma_{\vec{k}-\vec{p}}(\tau),
\end{equation}
where $\hat{O}_{X}$ is an operator acting on $\delta\sigma$. The solution to Eq.~(\ref{eeom1}) is the combination of the general solution of the homogeneous equation (the standard vacuum solution) and a particular solution of the inhomogeneous equation
\begin{equation}
Q_{X}(\tau,\vec{k})=Q_{X}^{(v)}(\tau,\vec{k})+Q_{X}^{(s)}(\tau,\vec{k}),
\end{equation}
where
\begin{equation}\label{ggf}
Q_{X}^{(s)}(\tau,\vec{k})=\int^{\tau} d\tau^{'}\,G_{k}(\tau,\tau^{'})J_{X}(\tau',\vec{k})\, ,
\end{equation}
and the Green's function is given by 
\begin{equation}\label{gff}
G_{k}(\tau,\tau^{'})\simeq\frac{1}{k^{3}\tau\tau^{'}}\left[k\tau'\cos(k\tau')-\sin(k\tau')\right],\quad\quad  -k\tau\ll 1.
\end{equation}
The power spectrum of $Q_{X}$ will therefore have two contributions
\begin{equation}
P_{Q_{X}}=P^{(v)}_{Q_{X}}+P^{(s)}_{Q_{X}}, \quad\quad \langle  Q_{X}(\vec{k}_{1})Q_{X}(\vec{k}_{2})\rangle=\delta^{(3)}(\vec{k}_{1}+\vec{k}_{2})P_{Q_{X}}(k_{1}),
\end{equation}
where the first power spectrum, $P^{(v)}_{Q_{X}}$, comes from the vacuum fluctuations of the fields and the second, $P^{(s)}_{Q_{X}}$, is formally given by
\begin{equation}
\langle Q_{X}^{(s)}(\tau,\vec{k}_{1})Q_{X}^{(s)}(\tau,\vec{k}_{2})  \rangle =\int^{\tau} d\tau_{1} G_{k_{1}}(\tau,\tau_{1})\int^{\tau} d\tau_{2} G_{k_{2}}(\tau,\tau_{2})\langle J_{X}(\tau_{1},\vec{k}_{1}) J_{X}(\tau_{2},\vec{k}_{2}) \rangle.
\end{equation}
The results for the sourced power spectra are presented in the following two sections.

\section{Production of gravity waves}
\label{sec:GW}

In the spatially flat gauge, $\delta g_{ij}|_{scalar}=0$, the metric has the form $g_{ij} = a^2 \left( \delta_{ij} + h_{ij} \right)$, where $h_{ij}$  (transverse and traceless) incorporates the tensor degrees of freedom. The canonically normalized tensor modes are \footnote{The  circular polarization vectors satisfy  $\vec{k}\cdot \vec{\epsilon}^{\;(\pm)}  ( \vec{k} ) = 0$, $\vec{k} \times \vec{\epsilon}^{\;(\pm)} ( \vec{k} ) = \mp i k \vec{\epsilon}^{\;(\pm)} ( \vec{k} )$,
$\vec{\epsilon}^{\;(\pm)} ( -\vec{k} ) = \vec{\epsilon}^{\;(\pm)} ( \vec{k})^*$, and are normalized according to $\vec{\epsilon}^{\,(\lambda)}( \vec{k} )^* 
\cdot \vec{\epsilon}^{\,(\lambda')} ( \vec{k} ) = \delta_{\lambda \lambda'}$. }
\begin{equation}
 \frac{M_p a}{2} h_{ij}(\tau,\vec{x}) =  \int \frac{d^3 k}{\left( 2 \pi \right)^{3/2}} \, {\rm e}^{i \vec{k} \cdot \vec{x}}  \, \sum_{ \lambda =  \pm } \, \Pi_{ij,\lambda} ( {\hat k} ) \,  \hat{h}_\lambda ( \tau,\vec{k} ) \;\;\;,\;\;\;
\Pi_{ij,\lambda} ( {\hat k} ) = \epsilon^{(\lambda)}_i ( {\hat k} )  \epsilon^{(\lambda)}_j ( {\hat k} ) ,
\label{formal-hc-def}
\end{equation}
and in momentum space they obey the equation 
\begin{equation}\label{src}
\left[ \partial_\tau^2 + \left( k^2 - \frac{a''}{a} \right) \right]  \hat{h}_\lambda ( \tau ,\, \vec{k} ) = J_{\lambda} ( \tau ,\, \vec{k} ).
\end{equation}
To leading order in slow-roll one finds
\begin{equation}\label{srct}
J_{\lambda} ( \tau ,\, \vec{k} )\simeq 2\,a\,\epsilon\left(\frac{\Lambda^{2}H^{2}}{M^{4}}\right)\Pi_{ij,\lambda} ( {\hat k} ) \int \frac{d^{3}p}{(2\pi)^{3/2}}p_{i}p_{j}\delta\sigma_{\vec{p}}(\tau)\delta\sigma_{\vec{k} - \vec{p}}(\tau).
\end{equation}
The derivation of the source term (\ref{srct}) is straightforward if one expands the Lagrangian to third order and identifies the interactions of the type $\sim \mathcal{O}(h\delta\sigma\delta\sigma)$. Using (\ref{ggf}) and the Green's function (\ref{gff}), one obtains (see also \cite{Biagetti:2013kwa}) for the sourced tensor power spectrum 
\begin{equation}
P_{\lambda}^{(s)}=\frac{1}{a^{2}(\tau)}\int \frac{d^{3}p}{(2\pi)^{3}} \,p^{4}\sin^{4}\theta\,\Big| \int_{-\infty}^{\tau} d\tau^{'}a(\tau^{'})G_{k}(\tau,\tau^{'})\delta\sigma_{p}(\tau^{'})\delta\sigma_{|\vec{k}-\vec{p}|}(\tau^{'})\Big|^{2}.
\end{equation}
Using new variables
\begin{equation}\label{newvar}
p\equiv\frac{ky}{c_{s}},\quad |\vec{k}-\vec{p}|\equiv\frac{kx}{c_{s}},\quad z=k\tau,
\end{equation}
and the expression for the spectator field mode-functions, Eq.~(\ref{sfmf}), one has
\begin{equation}\label{qu}
P_{\lambda}^{(s)}= \left(\frac{\epsilon \Lambda^{2}H^{2}}{M^{4}}\right)^{2}\frac{z^{2}B^{4}}{(2\pi)^{2}M_{P}^{4}\, k^{3}\,c_{s}^{2}}\int dx\,dy\, \frac{y^{2}}{x^{2}}\left[1-\left(\frac{c_{s}^{2}+y^{2}-x^{2}}{2c_{s}y}\right)^{2}\right]^{2}F^{2}(x,y,z),
\end{equation}
where $B$ is the normalization of the spectator field mode function, Eq.~(\ref{sfmf}), and
\begin{equation}
F^{2}(x,y,z)\equiv\Big| \int_{-\infty}^{z} \frac{dz}{zz^{'2}}\left(z^{'}\cos z^{'}-\sin z^{'}\right)\left(1+i\,yz^{'}\right)\left(1+i\,xz^{'}\right)e^{-iz'(x+y)} \Big|^{2}.
\end{equation}
The $(x,y)$ integral can be performed numerically, leading to the final result
\begin{equation}\label{tps}
P_{\lambda}^{(s)}\simeq \frac{0.1}{\pi^{2}}\,\left(\frac{H}{M_{P}}\right)^{4}\frac{1}{c_{s}^{6.2}}\frac{1}{k^{3}},
\end{equation}
where we also used the relation between the sound speed and the slow-roll parameter that is specific of our model, $c_{s}^{2}=\epsilon/3(n-1)$ (see Appendix~\ref{A}).

\section{Production of scalar perturbations}
\label{sec:zeta}

The inflaton and the spectator field fluctuations are only coupled through the metric perturbations. One derives the non-linear equation of motion for $\delta\phi$ and $\delta\sigma$ by integrating out the non-dynamical scalar degrees of freedom of the metric.~Major contributions to the source for the scalar field comes from cubic interactions which are quadratic in the spectator field~
\footnote{The inflaton and scalar fields already couple at quadratic level, which would correspond to a tree-level correction to the power spectrum of $\delta\phi$. The $\delta\sigma$$\delta\phi$ mixing terms in the second order Lagrangian vanish when the background energy density of $\sigma$ vanishes, namely when  $\rho_{\sigma}/(H^{2}M_{P}^{2} )\rightarrow 0$ (see Appendix~A). For finite $\rho_\sigma$, this mixing provides an additional 
source of $\delta \phi$. The amount of scalar perturbations that we compute from the $\delta \sigma^2 \rightarrow \delta \phi$ process is not proportional to this parameter, which leads us to conclude that  the extra terms $\delta \phi$ arising from the linear mixing can be disregarded at sufficiently small  $\rho_\sigma$. Nonetheless, to remind that such extra terms are present, and could be relevant in some range of parameters, we conservatively present  our results in Sec.~(\ref{sec:pheno}) as a lower bound on the sourced power spectrum (and consequently as an upper bound to the tensor-to-scalar ratio).}

The calculation is presented in Appendix~\ref{B}, here we report the final result 
\begin{equation}
v'' + \left( -\Delta- \frac{a''}{a} \right) v \simeq J_{v},
\end{equation}
\begin{eqnarray}\label{sscalar}
&&\frac{J_{v}}{a}\equiv -\frac{1}{\sqrt{2\epsilon}M_P}
\frac{\Lambda^{2}H^{2}}{M^{4}}\frac{1}{\mathcal{H}^{2}}\Biggl\{ 6\Delta^{-1}\Bigl[( \Delta \delta\sigma')^2 - \partial_{i}\partial_{j}\dsg' \partial_{i}\partial_{j}\dsg'+  \Delta \delta\sigma \Delta \delta\sigma'' - \partial_{i}\partial_{j}\dsg \partial_{i}\partial_{j}\delta\sigma'' \Bigl]  \nonumber\\&&\quad\quad -2 \dij\Bigl[\partial_{i}\dsg' \partial_{j}\dsg'+\partial_{i}\partial_{j}\dsg \Delta\dsg-\partial_{i}\partial_{k}\dsg\partial_{j}\partial_{k}\dsg -\dsg'' \partial_{i}\partial_{j}\dsg \Bigl]\nonumber\\  & &\quad\quad-4\Bigl[\delta\sigma''\Delta\dsg+2 \delta\sigma'\Delta\dsg'\Bigl]\Biggl\} -\frac{12}{\sqrt{2\epsilon}M_{P}}\frac{\Lambda^{2}H^{2}}{M^{4}}\partial_{i}\delta\sigma\left[-\frac{2}{\mathcal{H}}\partial_{i} \delta\sigma^{'}+\partial_{i} \delta\sigma\right],
\end{eqnarray}
where the canonically normalized field is $v\equiv a\delta\phi$ and $\mathcal{D}_{ij} = \delta_{ij} - 3\Delta^{-1}\partial_i\partial_j$.  \\
\indent The spectator field contribution to the scalar power-spectrum arising from $\delta\phi\delta\sigma\delta\sigma$ interactions can be computed in the same way as in the previous section. The source term (\ref{sscalar}) has several different contributions. For the sake of a quick comparison with the result of the previous section, one may begin with the last term in Eq.~(\ref{sscalar}), which gives a contribution to the sourced scalar power spectrum 
\begin{equation}\label{sourcp}
P_{\zeta}^{(s)}\supset\frac{H^{2}}{\dot{\phi}^{2}}\left(\frac{\Lambda^{2}H^{2}}{\epsilon M^{4}M_{P}^{4}}\right)^{2}\frac{1}{a^{2}(\tau)}\int \frac{d^{3}p}{(2\pi)^{3}} \left(\vec{p}\cdot(\vec{k}-\vec{p})\right)^{2}\Big| \int_{-\infty}^{\tau} d\tau^{'}a(\tau^{'})G_{k}(\tau,\tau^{'})\delta\sigma_{p}(\tau^{'})\delta\sigma_{|\vec{k}-\vec{p}|}(\tau^{'})\Big|^{2}
\end{equation}
and in the variables (\ref{newvar}) becomes
\begin{equation}
P_{\zeta}^{(s)}\supset \frac{H^{2}}{\dot{\phi}^{2}}\left(\frac{\Lambda^{2}H^{2}}{\sqrt{\epsilon} M^{4}M_{P}}\right)^{2}\frac{B^{4} z^{2}}{(2\pi)^{2}c_{s}^{2}k^{3}}\int dx\,dy\, \frac{y^{2}}{x^{2}}\left[1-\left(\frac{c_{s}^{2}+y^{2}-x^{2}}{2y^{2}}\right)\right]^{2}F^{2}(x,y,z).
\end{equation}
After integration one finds
\begin{equation}\label{rrr}
P_{\zeta}^{(s)}\supset \frac{10}{\pi^{2}}\,\left(\frac{H}{M_{P}}\right)^{4}\frac{1}{\epsilon^{4}c_{s}^{6.4}}\frac{1}{k^{3}}.
\end{equation}
Notice that for this specific contribution the power spectrum of curvature fluctuations has a stronger enhancement compared to the tensor power spectrum (\ref{tps}). A stronger enhancement of the scalars than the one found for the tensors can also be expected from the other terms in the source (\ref{sscalar}). With the change of variables introduced above in Eq.~(\ref{newvar}), it is easy to derive a quick estimate of the contributions from the different terms in $J_{v}$ to the scalar power spectrum. Notice, first of all, that every temporal derivative is accompanied  by an inverse power of $\mathcal{H}$
\begin{eqnarray}
\delta\sigma^{'}(\tau,p)\sim  p c_{s} \delta\sigma(\tau,p)\sim k y,\quad\quad \delta\sigma^{''}(\tau,p)\sim p^{2} c_{s}^{2} \delta\sigma \sim k^{2} y^{2},\nonumber\\
\mathcal{H}=aH\sim \frac{1}{\tau} \sim\frac{z}{k},\quad\quad\quad\quad\quad\quad\quad\quad
\end{eqnarray}
extra temporal derivative acting on $\delta\sigma$ fluctuations are therefore not expected to modify the power spectrum overall amplitude w.r.t. to the contribution in (\ref{sourcp}). On the other hand, each spatial derivative brings (going to Fourier space) an enhancement with $c_{s}$  
\begin{equation}
\partial_{i}\delta\sigma\sim \frac{k}{c_{s}}.
\end{equation}
Then one can write
\begin{equation}\label{recipe}
P_{\zeta}^{(s)JJ'}\sim \left(\frac{H^{4}}{M_{P}^{4}}\right) \left(\frac{1}{\epsilon^{4}}\right) \left(\frac{1}{k^{3}}\right) \left(\mathcal{O}_{JJ'}\right)^{2},\quad\quad \mathcal{O}_{JJ'}\equiv c_{s}^{-1 \times \,(\# space\,der)},
\end{equation}
where indices $JJ'$ refer to the combination any two terms of $J_{v}$ (or of a single term with itself) in the two-vertex loop diagram. This estimate carries an uncertainty in the power of $c_{s}$ of the final result for the power spectrum, it nevertheless can be very handy for a quick evaluation of the magnitude of the other contributions in Eq.~(\ref{sscalar}). By looking at (\ref{sscalar}) and using (\ref{recipe}), what one finds is that the contribution in Eq.~(\ref{rrr}) is likely the smallest one derived from $J_{v}$, the others being of the same magnitude or $c_{s}^{-2}$ enhanced w.r.t. (\ref{rrr}) (from diagrams with two vertices associated to interactions carrying respectively 2 spatial derivatives and 4 spatial derivatives), or $c_{s}^{-4}$ enhanced (from diagrams with two vertices both carrying 4 spatial derivatives).\\

The effect of these sourced power spectra on the tensor-to-scalar ratio is detailed in the next section. We underscore here that, whenever the $\sigma$-sourced power spectra (scalar and tensor or scalar only) become leading compared to the vacuum contributions, one crucial characteristic of the theory emerges: the knowledge of $r$ and of the scalar spectrum does not necessarily determine $H$; a clear-cut correspondence occurs only in a subset of inflationary theories.

\section{Summary of the results and phenomenology}
\label{sec:pheno}

The final results for the total scalar and tensor power spectra read
\begin{eqnarray}\label{efr}
 P_{\lambda}=P_{\lambda}^{(v)}+P_{\lambda}^{(s)},\\\label{efra}
 P_{\zeta}=P_{\zeta}^{(v)}+P_{\zeta}^{(s)},
\end{eqnarray}
where from the previous sections the sourced contributions are
\begin{eqnarray}\label{frr}
P_{\lambda}^{(v)}=  \frac{2H^{2}}{M_{P}^{2}k^{3}} ,\quad\quad\quad\quad\quad\quad\quad\quad\quad   P_{\lambda}^{(s)}\simeq \frac{0.1}{\pi^{2}} \frac{H^{4}}{M_{P}^{4}}\frac{1}{c_{s}^{6}k^{3}}  ,\nonumber\\\label{abov}
P_{\zeta}^{(v)}= \frac{H^{2}}{4\epsilon M_{P}^{2}k^{3}}    ,\quad\quad\quad\quad\quad\quad\quad\quad P_{\zeta}^{(s)}\gtrsim    \frac{H^{4}}{M_{P}^{4}}\frac{1}{\pi^{2}\epsilon^{4}c_{s}^{10}k^{3}}.
\end{eqnarray}
For the sake of clarity, we have chosen to disregard the decimal figures in the power of $c_{s}$ in Eqs.(\ref{tps}) and (\ref{rrr}), since they cannot affect the comments and conclusions that are offered in this and the next section.

The scalar power spectrum must satisfy the normalization 
\begin{equation}
2.5 \cdot 10^{-9} = {\cal P}_\zeta \equiv \frac{k^3}{2 \pi^2} P_\zeta \ga  \frac{H^2}{ 8 \pi^2 \epsilon M_p^2} \left[ 1 + {\mathcal{O} } \left( \frac{\alpha}{\epsilon^3 c_s^4} \right)  \right]  \,,\quad\quad \alpha\equiv\frac{H^{2}}{\pi^{2}M_{P}^{2}c_{s}^{6}} 
\label{Pz-m}
\end{equation} 
and the tensor-to-scalar ratio is 
\begin{equation}
r\equiv \frac{\sum_{\lambda} P_{\lambda}}{P_{\zeta}}\lesssim 16 \epsilon \,\left[\frac{1+\alpha}{1+\mathcal{O}\left(\frac{\alpha}{\epsilon^{3} c_{s}^{4}}\right)}\right].
\label{r-result-m}
\end{equation}
The inequality in (\ref{Pz-m}) and (\ref{r-result-m}) is to account for the possible contributions to the sourced scalar power spectrum from quadratic interactions between the inflaton and the spectator field (these are suppressed at small  $\rho_\sigma$, see the remark in footnote\text{$^7$}). In the following, we work in a regime where these contributions are negligible, and promote $r$ to its upper bound. In this way we can get an estimate for how large $r$ can be in this model.

We are now ready to discuss the estimate (\ref{r-result-m}) in various regimes for $c_{s}$. The first regime is that in which the vacuum scalar modes dominate over the sourced one (and, therefore, also the vacuum tensor modes dominate over the sourced one). In this regime, 
\begin{equation}
\alpha \ll \epsilon^3 \, c_s^4 \;\;\Rightarrow\;\; 
r = 16 \, \epsilon \;, 
\label{r1-m}
\end{equation}
and so the standard relation for $r$ is recovered. We see from (\ref{Pz-m}) that $(H^2/M_p^2 )\simeq 2 \cdot 10^{-7} \epsilon$, so that the condition in (\ref{r1-m}) rewrites $c_s \gg {\mathcal{O} } \left( 10^{-1} \, \epsilon^{-1/5} \right)$. 

Let us then discuss the opposite regime, in which the sourced scalar modes dominate, and $r$ differs from the standard value. In this regime, the normalization (\ref{Pz-m}) enforces 
\begin{equation}
\alpha \gg \epsilon^3 \, c_s^4 \;\;\Rightarrow\;\;  \frac{H}{M_p} \sim 0.04 \, \epsilon \, c_s^{5/2}  \,,
\label{H2-m}
\end{equation} 
so that the parameter $\alpha$ evaluates to $\alpha \sim  10^{-4} \, (\epsilon^2/c_s)$, indicating that this regime holds for  $c_s \ll {\mathcal{O} } \left( 10^{-1} \, \epsilon^{-1/5} \right)$ (this is the complementary regime to the one of (\ref{r1-m})). Inserting the estimate for $\alpha$ into (\ref{r-result-m}), we obtain 
\begin{equation}
r \sim  10^5 \, \epsilon^2 c_s^5 \left[ 1 +   10^{-4} \, \frac{\epsilon^2}{c_s} \right] \,. 
\label{r2-m}
\end{equation}

We disregard the regime in which the sourced tensor modes dominate (namely, in which the second term in the square parenthesis of this last expression is greater than one), as this regime requires the very small sound speed 
$c_s \la {\mathcal{O} } \left( 10^{-4} \epsilon^2 \right)$. We are therefore left with 
\begin{equation}
10^{-4} \epsilon^2 \ll c_s \ll \frac{1}{10 \, \epsilon^{1/5} } \;\;\Rightarrow\;\; 
\frac{H}{M_p} \sim {\mathcal{O} } \left( 10^{-2} \, \epsilon \, c_s^{5/2} \right) \;\;,\;\; 
r \sim {\mathcal{O} } \left( 10^5 \epsilon^2 c_s^5 \right) \;, 
\label{final-m}
\end{equation}
so that, if $c_s$ is close to the upper bound, $r$ may be as large as ${\mathcal{O} } \left( \epsilon \right)$. 

We conclude this discussion by noting that (\ref{final-m}) assumes that the sourced scalar modes dominate over the vacuum ones. This calls for an investigation on whether the scalar perturbations in this regime may be compatible with the observed Gaussianity of the CMB� \cite{Ade:2013ydc}. Indeed, Ref.�\cite{Barnaby:2010vf} studied the sourced scalar modes from a $\delta A \delta A \rightarrow \zeta$ process in a different model, which resulted in the requirement that $\zeta_{\rm sourced}$ be subdominant in that model. This all goes to show that the study of non-Gaussianity of the scalar perturbations is the crucial next step to unravel the complete phenomenology of the model considered here, and to ultimately place constraints on the sound speed of $\sigma$ (see also \cite{Baumann:2014cja}). We leave this to future work.

\section{Conclusions}

The search for imprints of primordial gravitational waves has been at the center of numerous experimental efforts, recently contributed by the first observation of B-mode polarizations in the CMB. Their importance lies in the fact that gravitational waves generated during inflation carry precious information. In inflation, the amplitude of gw is often directly related to $H$ and the implication of a large enough primordial tensor signal can place tight constraints on models, likely ruling out most small-field models. These statements (which can be summarized as Eqs.~(\ref{one}) and (\ref{two})) normally rely on the assumption that primordial gw originate from the vacuum fluctuations of the tensor modes of the metric, so they do not apply to all inflationary models. If alternative production mechanisms, occurring during inflation or at later times, were responsible for the observed B-mode fluctuations, the implications of an observed primordial gw signal for inflation could be entirely different from those one is lead to when assuming generation from vacuum fluctuations as the culprit. \\

It is therefore crucial to establish to what extent and under which conditions (\ref{one}) and (\ref{two}) represent a prediction of inflation. \\

Gravitational waves can also be sourced at second order by quantum fluctuations of fields during inflation (they are also generated at second order already by scalar curvature perturbations alone \cite{gwg1}). Predictions are model dependent (e.g. depending on the spin of the fields involved, the nature of their couplings to the inflaton and the details of their dynamics) and, in order to assess whether a given source of gravity waves enhances the observable tensor-to-scalar ratio, one needs to compute the scalar perturbations that this source will also produce  \cite{Barnaby:2012xt}. \\

\indent Recently, a model has been put forward \cite{Biagetti:2013kwa} where inflation occurs in the presence of a spectator field, $\sigma$, not responsible for driving the expansion nor for directly providing a substantial contribution to the primordial curvature fluctuations. The field $\sigma$ has a sound speed $c_{s}< 1$ and it sources second order tensor fluctuations. The smallness of the sound speed can counteract the suppression that one would normally expect at non-linear order. The dynamics of the fluctuations $\delta\sigma$ in \cite{Biagetti:2013kwa} is described by the effective Lagrangian (\ref{chico}) and the small values of the sound speed can result in an enhancement of the tensor power spectrum compared to the contribution from vacuum fluctuations only.\\

In this work we have proposed and studied a covariant formulation of \cite{Biagetti:2013kwa}. We have considered an inflaton, $\phi$, with a canonical Lagrangian and an auxiliary field, $\sigma$, with a small sound speed $c_{s}$. The fields $\phi$ and $\sigma$ are minimally coupled with one another, and $\sigma$ has a non-minimal derivative coupling with gravity. Spectator field-sourced contributions to the power spectrum of curvature fluctuations, $\zeta$, and to the power spectrum of gravitational waves arise in particular from interactions of $\sigma$ with the inflaton $\phi$ and with the tensor fluctuations of the metric. We have studied this model in some details, shown e.g. how the non-minimal coupling is a desirable feature in the search of constant energy density solutions and how the sign of this term is dictated by the requirement on the absence of ghosts.\\

We departed from the approach in \cite{Biagetti:2013kwa} in that we wrote down a $P(X_{\sigma})$ Lagrangian for $\sigma$, as opposed to an effective Lagrangian for the $\sigma$ fluctuations. Our findings are then limited by this initial assumption but allow for a fairly general (general to, say, the same extent that a $P(X)$ inflationary  model is general) no-go statement on the possibility of enhancing the tensor to scalar ratio via ``slow" spectator fields.  We should also stress that, despite the $\sigma$ field ``spectator" nature, its presence is enough to break, via $c_s$, the clear-cut correspondence between the knowledge of $r, P_{\zeta}$ and that of the all-important energy scale of inflation for curvature fluctuations. 

It would be interesting to study if alternative covariant formulations of \cite{Biagetti:2013kwa} can be obtained where the scalar density production from $\sigma$ is more suppressed, so that a greater value of $r$ could result. It would also be a worthwhile endeavor to investigate the precise role the $\sigma$ field plays in determining the amplitude and shape of the non-Gaussian signal. We leave this to future work.

\acknowledgments

It is a pleasure to thank A.~Riotto for collaboration at the early stages of this project and for insightful conversations. MF is grateful to A.~J.~Tolley for illuminating discussions.\\
M.B. acknowledges support by the Swiss National Science Foundation. The work of MP was supported in part by DOE grant DE-SC0011842 at the University of Minnesota. The work of ED was supported by the Department of Energy at ASU. The work of MF was supported in part by grants DE-SC0010600 and NSF PHY-1068380. \\

\appendix
\numberwithin{equation}{section}

\section{Full linear perturbations analysis}
\label{A}

In this section we present the derivation of the second order Lagrangian for the inflaton and for the spectator field, including metric perturbations. The metric has the form
\ba
g_{00}&=&-a^{2}\left(1-2\Phi\right),\nonumber\\
g_{0i}&=&a^{2}\left(B_{i}+\partial_{i} B\right),\nonumber\\
g_{ij}&=&-a^{2}\left[\left(1+2\psi\right)\delta_{ij}+2\partial_{i}\partial_{j}E+\partial_{i}E_{j}+\partial_{j}E_{i}+h_{ij}\right],
\ea
where $B_{i}$ and $E_{i}$ are vector modes (transverse) and $h_{ij}$ is a traceless and transverse tensor. When studying scalar fluctuations at linear order, vector and tensor fluctuations can be ignored (scalar, vector and tensor fluctuations decouple at linear order). It is always possible to choose a gauge in which $\psi=E=0$, so one is left with $\Phi$ ad $B$ fluctuations only.\\

It is convenient to introduce rescaled hat-variables, related to those of the inflaton and of the spectator field by 
\begin{equation}
\delta \phi = \frac{\delta {\hat \phi}}{a} \;\;,\;\; \quad\quad
\delta \sigma = \frac{M^2 a}{\sqrt{12 \left( n - 1 \right) } \Lambda a'} \delta {\hat \sigma} 
\end{equation} 
and to define  $X^T = \left( \delta {\hat \phi } ,   \delta  {\hat \sigma } \right) \;,\; N^T = \left( \Phi, B \right)$. One then obtains 
\begin{eqnarray} 
S = \int d \tau d^3 k \Bigg[ X^{'\dagger} A X' + \left( X^{\dagger '} B X + {\rm h. c.} \right) +  X^\dagger C X + \left( N^\dagger D X' + {\rm h.c.} \right) \nonumber\\
 +  \left( N^\dagger E X + {\rm h.c.} \right) + N^\dagger F N \Bigg] 
\;\;,
\end{eqnarray} 
with 
\begin{eqnarray}
A &=& \left( \begin{array}{cc} 
\frac{1}{2} & 0 \\ 0 & \frac{1}{2} 
\end{array} \right) ,\nonumber\\ 
B &=& a \left( \begin{array}{cc} 
- \frac{ H}{2} & 0 \\ 0 &   - \frac{H}{2} + \frac{\phi'^2}{4 a^2 M_p^2 H \left( 1 - \frac{n \rho_\sigma}{3 M_p^2 \left( n - 1 \right) H^2 } \right) }  
\end{array} \right), \nonumber\\ 
C & = & \frac{a^2}{2} \left( \begin{array}{cc} 
- p^2 + H^2 - V'' & 0 \\ 
0 & H^2 - \frac{\phi'^2}{a^2 M_p^2} \, \frac{1}{1-\frac{n \rho_\sigma}{3 M_p^2 \left( n - 1 \right) H^2} } 
\left[ 1 - \frac{\phi'^2}{4 M_p^2 a^2 H^2} \frac{1}{1-\frac{n \rho_\sigma}{3 M_p^2 \left( n - 1 \right) H^2} } + \frac{p^2}{6 \left( n - 1 \right) H^2} \right] 
\end{array} \right), \nonumber\\ 
D &=& a^2 \left( \begin{array}{cc} 
- \frac{1}{2} \frac{\phi'}{a} & - \frac{  n - 2 }{ \sqrt{n+1}} \, \frac{\sqrt{n}}{\sqrt{n-1}} \, \sqrt{-\rho_\sigma} \\ 0 & - a \frac{ p^2 \sqrt{n}  \sqrt{-\rho_\sigma}}{3 \sqrt{n-1} \sqrt{n+1} H} 
\end{array} \right) ,\nonumber\\ 
E & = &  \left( \begin{array}{cc} 
\frac{a^3}{2} \left( - V' + H \frac{\phi'}{a} \right) & - \frac{a^3}{\sqrt{n-1}} \frac{\sqrt{n}}{\sqrt{n+1}} \sqrt{-\rho_\sigma} 
\frac{1}{H} \left[ \frac{p^2}{3} + H^2 \left( n - 2 \right) 
 \left( 1 - \frac{\phi'^2}{2 M_p^2 H^2 a^2}   
 \frac{1}{1-\frac{n \rho_\sigma}{3 M_p^2 \left( n - 1 \right) H^2} } \right) \right]  \\ 
 - \frac{a^4}{2} p^2 \frac{\phi'}{a} &  -  \frac{a^4}{\sqrt{n-1}} \frac{\sqrt{n}}{\sqrt{n+1}} \sqrt{-\rho_\sigma} 
  \frac{p^2}{3}  \left( 1 - \frac{\phi'^2}{2 M_p^2 H^2 a^2}   
 \frac{1}{1-\frac{n \rho_\sigma}{3 M_p^2 \left( n - 1 \right) H^2} } \right)     
\end{array} \right), \nonumber\\
F & = &  \left( \begin{array}{cc} 
a^4 \left[  - 3 H^2 M_p^2 + \frac{\phi'^2}{2 a^2} - \frac{ \left(2 n - 7 \right) n \rho_\sigma}{n+1} \right]   
&  \frac{a^5 p^2}{H} \left(  M_p^2 H^2 - \frac{n \rho_\sigma}{n+1}  \right)   \\ 
 \frac{a^5 p^2}{H} \left(  M_p^2 H^2 - \frac{n \rho_\sigma}{n+1}  \right) &  0  
\end{array} \right). \nonumber\\ 
\end{eqnarray}

\noindent Integrating out the nondynamical fields one finds 
\begin{eqnarray}\label{matA}
&& S = \int d \tau d^3 k \left[ X^{'\dagger} \,  K \,  X' + 
\left( X^{'\dagger} \, \Theta \,  X + {\rm h.c.} \right)
+ X^\dagger \left( C - E^\dagger F^{-1} E \right) X \right] ,\nonumber\\ 
&&  \nonumber\\
&&  K  \equiv  A - D^\dagger F^{-1} D , \nonumber\\ 
&& \Theta \equiv  \left( B - D^\dagger F^{-1} E \right) ,\nonumber\\ 
&& \Omega \equiv   \left( C - E^\dagger F^{-1} E \right) .
\end{eqnarray}

\noindent Defining the  parameters\footnote{Notice the distinction between $\epsilon$ and $\epsilon_\phi$, although these two quantities turn out to be the same at leading order in slow roll. }
\begin{equation}
\epsilon \equiv \frac{\dot{\phi}^2}{2 Mp^2 H^2} \;,\; \epsilon_\phi \equiv \frac{M_p^2}{2} \frac{V^{'2}}{V^2} \;,\; 
\eta = M_p^2 \,  \frac{V''}{V} \;,\; \epsilon_\sigma \equiv - \frac{\rho_\sigma}{M_p^2 H^2} ,
\end{equation} 
the matrices in Eq.~(\ref{matA}) are given by (all these expressions are exact)
\begin{eqnarray} 
K_{11} &=& \frac{1}{2}\,, \nonumber\\ 
K_{12} &=& K_{21} = \frac{-1}{3 \sqrt{2}} \, \frac{\sqrt{\frac{\epsilon}{n-1} \, \frac{n \epsilon_\sigma}{n+1}}}{1+ \frac{n \epsilon_\sigma}{n+1}} \,,\nonumber\\ 
K_{22} &=& \frac{1}{2} - \frac{6 n-9-\epsilon+\left(4n-5\right) \frac{n \epsilon_\sigma}{n+1}}{9 \left( n-1\right) \left( 1 + \frac{n}{n+1} \epsilon_\sigma \right)^2} \, \frac{n \epsilon_\sigma}{n+1} 
\end{eqnarray} 
and
\begin{eqnarray}
\frac{\Theta_{11}}{a} & = & - \frac{H}{2} \, \left( 1 + \frac{\epsilon}{1+\frac{n \epsilon_\sigma}{n+1}} \right)\,, \nonumber\\ 
\frac{\Theta_{12}}{a} & = & K_{12} \, H \left( 1 - \frac{\epsilon}{1+\frac{n \epsilon_\sigma}{3\left(n-1\right)} } \right) \,,\nonumber\\ 
\frac{\Theta_{21}}{a} & = & - K_{12} \, H \left[ 3 \sqrt{\frac{\epsilon_\phi}{\epsilon}} \left( 1 + \frac{\epsilon_\sigma-\epsilon}{3} \right) - \frac{3 n - 4 - \epsilon + \frac{n^2 \epsilon_\sigma}{n+1} }{1+ \frac{n \epsilon_\sigma}{n+1}} \right]  \,, \nonumber\\ 
\frac{\Theta_{22}}{a} & = &  - \frac{p^2}{9 H} \frac{n \epsilon_\sigma}{\left( n-1 \right) \left( n+1 \right) \left( 1 + \frac{n \epsilon_\sigma}{n+1} \right) } - \frac{1}{2} H \left( 1 - \epsilon \right) \nonumber\\ 
& & \!\!\!\!\!\!\!\!  \!\!\!\!\!\!\!\!    - \frac{H \frac{n \epsilon_\sigma}{n+1} }{54 \left( n - 1 \right) \left( 1 + \frac{n \epsilon_\sigma}{n+1} \right)^2 \left( 1 + \frac{n \epsilon_\sigma}{3 \left( n - 1 \right)} \right)} 
\Bigg[ - 54 + 36 n + 57 \epsilon - 27 n \epsilon + 6 \epsilon^2\nonumber\\ 
& &  \!\!\!\!\!\!\!\!  + \frac{n \epsilon_\sigma}{n+1} \left( 12 \left( 4 n - 1 \right) - 12 \frac{n^2}{n-1} + 52 \epsilon - 2 \frac{3 n^2-1}{n-1} \epsilon - 2 \frac{n \epsilon_\sigma}{n-1} + 8 n \epsilon_\sigma + 9 n \epsilon \epsilon_\sigma \right) \Bigg] 
\nonumber\\ 
\end{eqnarray} 
and
\begin{eqnarray}
\frac{\Omega_{11}}{a^2} & = & - \frac{p^2}{2} + \frac{H^2}{2} \left[ 1 + \frac{\left(2n-5\right)\epsilon}{1+\frac{n \epsilon_\sigma}{n+1}} - \frac{\left( 2 n - 4 - \epsilon \right) \epsilon}{\left( 1 + \frac{n \epsilon_\sigma}{n+1} \right)^2} 
+ \frac{6 \sqrt{\epsilon \epsilon_\phi} \left( 1 + \frac{\epsilon_\sigma-\epsilon}{3} \right) }{1+\frac{n}{n+1}\epsilon_\sigma} \right] \nonumber\\ 
& & + \frac{3}{2} H^2 \left(  1 + \frac{\epsilon_\sigma-\epsilon}{3} \right) \eta \,,\nonumber\\ 
\frac{\Omega_{12}}{a^2} & = & \frac{\Omega_{21}}{a^2}  =   K_{12} p^2 + 3 K_{12} H^2 \Bigg[ 
\frac{\left( 1 - \epsilon + \frac{n \epsilon_\sigma}{3\left(n-1\right)} \right) \left( 3 \left( n - 1 \right) - 1 - \epsilon + \frac{n^2}{n+1} \epsilon_\sigma \right) }{3 \left( 1 + \frac{n \epsilon_\sigma}{n+1} \right) \left( 1 + \frac{n \epsilon_\sigma}{3 \left( n - 1 \right)} \right) }  \, \nonumber\\ 
& & \quad\quad  \quad\quad  \quad\quad  - \sqrt{\frac{\epsilon_\sigma}{\epsilon}} \, \frac{ \left( 1 + \frac{\epsilon_\sigma - \epsilon}{3} \right) \left( 1 - \epsilon + \frac{n \epsilon_\sigma}{3 \left( n - 1 \right)} \right) }{1+\frac{n \epsilon_\sigma}{3 \left( n - 1 \right)  } } \Bigg] \,,\nonumber\\ 
\frac{\Omega_{22}}{a^2} & = & - p^2 \; \frac{4 \frac{n \epsilon_\sigma}{n+1} \left( 1 + \frac{n \epsilon_\sigma}{3 \left( n - 1 \right)} \right) + 3 \epsilon \left( 1 - \frac{n \epsilon_\sigma}{3 \left( n + 1 \right) } \right)}{18 \left( n - 1 \right) 
\left( 1 + \frac{n \epsilon_\sigma}{n+1} \right) \left( 1 + \frac{n \epsilon_\sigma}{3 \left( n - 1 \right) } \right) } \nonumber\\ 
& & + H^2 \; \frac{\left( 1 - \epsilon + \frac{n \epsilon_\sigma}{3 \left( n - 1 \right)} \right)^2 \left( 9 + \frac{2 n }{n^2-1} \left( 3 n + \epsilon \right) \epsilon_\sigma + \frac{n^2}{n^2-1} \epsilon_\sigma^2 \right)}{18 
\left( 1 + \frac{n \epsilon_\sigma}{n+1} \right)^2 \left( 1 + \frac{n \epsilon_\sigma}{3 \left( n - 1 \right) } \right)^2 }  .
\end{eqnarray} 
The two fields decouple for $\epsilon_\sigma = 0$
\begin{eqnarray}
K & = &  \frac{1}{2} \left( \begin{array}{cc} 1 & 0 \\ 0 & 1 \end{array} \right)\,, \\ 
\Theta & = & - \frac{a H}{2}  \left( \begin{array}{cc} 1+\epsilon & 0 \\ 0 & 1-\epsilon \end{array} \right) \,,\nonumber\\ 
\Omega & = & - \frac{a^2 }{2}  \left( \begin{array}{cc} 
p^2 + H^2 \left( -1 - 6 \sqrt{\epsilon_\phi \epsilon} + \epsilon + 2 \sqrt{\epsilon_\phi } \epsilon^{3/2}
- \epsilon^2 + 3 \eta - \epsilon \eta \right) 
 & 0 \\ 0 &  \frac{\epsilon p^2}{3 \left( n - 1 \right)} - \left( 1 - \epsilon \right)^2 H^2   \end{array} \right) \nonumber
\end{eqnarray}
and the $22$ elements reproduces to leading order in slow-roll the quadratic Lagrangian for $\sigma$ in Sec.~(\ref{cov}). \\

\noindent The quadratic Lagrangian for the tensor modes is
\begin{equation}
\mathcal{S}_{\gamma^{2}}=\frac{M_{P}^{2}}{4}\int\,d^{3}x\, d\tau\,a^{2}\left(1+q\right)\left[h_{\lambda}^{'2}-\left(\frac{1-q}{1+q}\right)\left(\partial_{i}h_{\lambda}\right)^{2}\right]
\end{equation}
where
\begin{equation}
q\equiv \left(\frac{n}{n+1}\right)\left(\frac{-\rho_{\sigma}}{3H^{2}M_{P}^{2}}\right).
\end{equation}
Notice that in the regime where $q\ll 1$ one recovers the Lagrangian for simple single-field slow-roll inflation, with mode-functions given by
\begin{equation}
h_{\lambda}(\tau,k)=\frac{H}{M_{P}k^{3/2}}(1+ik\tau)e^{-ik\tau}.
\end{equation}

\section{Non-linear evolution equation for inflaton fluctuations}
\label{B}

In this section, more details are provided for the calculation of the source of scalar fluctuations arising from to third order interactions of the type $\sim \delta\phi\delta\sigma\delta\sigma$. One first derives the expressions of the $\delta g_{00}$ and $\delta g_{0i}$ modes of the metric in terms of the $\phi$ and $\sigma$ perturbations
\begin{equation}
\delta g_{00}=\delta g_{00}\left[\delta\phi,\delta\sigma\right]\,,  \quad\quad\quad \delta g_{0i}=\delta g_{0i}\left[\delta\phi,\delta\sigma\right].
\end{equation} 
Working in spatially flat gauge, the perturbed metric has the form
\begin{equation}
ds^2 = a^2 \left( \tau \right) \left[ - \left( 1 + 2 \Phi \right) d \tau^2 + 2 \partial_i B d \tau d x^i + \delta_{ij} d x^i d x^j \right],
\end{equation}
which gives
\begin{eqnarray}
G^0_0 & = &  - \frac{3 a'^2}{a^4} + \left\{ 
\frac{6 a'^2}{a^4 } \Phi + \frac{2 a'}{a^3}  \partial_k \partial_k B \right\}, \nonumber\\
G^0_i & = & - 2 \frac{a'}{a^3} \partial_i \Phi ,\nonumber\\
G^i_j & = & \left[ \frac{a'^2}{a^4} - 2 \frac{a''}{a^3} \right] \delta^i_j + 
\Bigg\{ \left[ - 2 \frac{a'^2}{a^4} \Phi + \frac{4 a''}{a^3} \Phi 
+ 2 \frac{a'}{a^3} \Phi' + \partial_k \partial_k \left( \frac{2 a'}{a^3} B + \frac{\Phi}{a^2} + \frac{B'}{a^2} \right) \right] \delta_{ij}
 \nonumber\\
& & \quad\quad\quad\quad \quad\quad\quad\quad 
+ \partial_i \partial_j \left( - \frac{2 a'}{a^3} B - \frac{\Phi}{a^2} - \frac{B'}{a^2} \right) \Bigg\}.
\end{eqnarray} 
The energy-momentum of the inflaton field is given by
\begin{eqnarray}
\left( T_{\phi} \right)^0_0 & = & - \frac{\phi'^2}{2 a^2} - V + \left( - V_{,\phi} \delta \phi + \frac{\phi'^2}{a^2} \Phi - \frac{\phi'}{a^2} \delta \phi' \right) ,\nonumber\\ 
\left( T_{\phi} \right)^0_i & = & - \frac{\phi'}{a^2} \partial_i \delta \phi, \nonumber\\ 
\left( T_{\phi} \right)^i_j & = & \delta^i_j \left[ \frac{\phi'^2}{2 a^2} - V +  \left( - V_{,\phi} \delta \phi - \frac{\phi'^2}{a^2} \Phi + \frac{\phi'}{a^2} \delta \phi' \right) \right] .
\end{eqnarray} 
Hamiltonian and momentum constraints read
\begin{eqnarray}\label{firsteq}
&& 6 M_p^2 {\cal H}^2 \Phi - \phi'^2 \Phi + \phi' \delta \phi' + a^2 V_{,\phi} \delta \phi + 2 M_p^2 {\cal H} \Delta B = a^2 
\left( T_\sigma \right)^0_0\,, \\\label{secondeq} 
&& - 2 M_p^2 {\cal H} \partial_i \Phi + \phi' \partial_i \delta \phi =  a^2 \left( T_\sigma \right)^0_i \,,
\end{eqnarray} 
where $T_{\sigma}$ is the energy momentum tensor for the spectator field.
From Eq.~(\ref{secondeq}) we have 
\begin{equation}
\Phi = \frac{1}{2 M_p^2 {\cal H}} \left[ \phi' \delta \phi - a^2 \Delta^{-1} \partial_i \left( T_\sigma \right)^0_i \right] \,;
\end{equation} 
inserting this in (\ref{firsteq}) we obtain
\begin{eqnarray}
\Delta B & = &  \frac{1}{2 M_p^2 {\cal H } }  \Bigg\{ - 3 {\cal H} \left( 1 - \frac{\phi'^2}{6 {\cal H}^2 M_p^2} \right) 
 \left[ \phi' \delta \phi - a^2 \Delta^{-1} \partial_i \left( T_\sigma \right)^0_i \right] \nonumber\\ 
 & & \quad\quad\quad\quad \quad - \phi' \delta \phi' - a^2 V_{,\phi} \delta \phi + a^2 \left( T_\sigma \right)^0_0 \Bigg\} \,.
\end{eqnarray} 
One writes the $ij$ Einstein equations in the form
\begin{equation}
a^2 M_p^2 G^i_j - a^2 \left( T_{\phi} \right)^i_j = a^2 \left( T_\sigma \right)^i_j \,,
\end{equation} 
obtaining
\begin{eqnarray} 
&& \delta_{ij} \left[ {\cal E}_1 + \Delta {\cal E}_2 \right] - \partial_i \partial_j {\cal E}_2 = a^2 \left( T_\sigma \right)^i_j \,,\nonumber\\ 
&& {\cal E}_1 \equiv 2 M_p^2 {\cal H} \Phi' + \left( 6 {\cal H}^2 M_p^2 - \phi'^2 \right) \Phi - \phi' \delta \phi' + a^2 V_{,\phi} \delta \phi \,,\nonumber\\ 
&& {\cal E}_2 \equiv M_p^2 \left( B' + 2 {\cal H} B + \Phi \right)\, .
\end{eqnarray} 
Notice that the spatial Einstein equations can be decoupled into the two scalar components
\begin{eqnarray}
{\cal E}_1 & = & a^2 \Delta^{-1} \partial_i \partial_j \left( T_\sigma \right)^i_j \,, \nonumber\\ 
\Delta {\cal E}_2 & = & \frac{a^2}{2} \left( T_\sigma \right)^i_i - \frac{3}{2} a^2 \Delta^{-1} \partial_i \partial_j 
 \left( T_\sigma \right)^i_j .
\end{eqnarray} 
Inserting the solution for $\Phi$, the equation for ${\cal E}_1$ becomes
\begin{equation}
\Delta^{-1} \partial_i \left\{ \left( \partial_\tau + 4 {\cal H} \right) \left( T_\sigma \right)^0_i + \partial_j \left( T_\sigma \right)^i_j \right\} = 0 \,,
\end{equation} 
which one further rewrites as 
\begin{equation}
\left( \partial_\tau + 6 {\cal H} \right) \left( T_\sigma \right)^{0i} + \partial_j \left( T_\sigma \right)^{ji} = \nabla_\mu 
\left( T_\sigma \right)^{\mu i}  = 0 .
\end{equation} 
Inserting the solutions for $\Phi$ and $B$, the equation for ${\cal E}_2$ becomes the master equation 
\begin{eqnarray}\label{mast}
v''-\left( \Delta + \frac{z''}{z} \right) v &=& a \Bigg\{ \left[ \frac{3 {\cal H}}{\phi'} - \frac{\phi'}{2 {\cal H} M_p^2} \right] {\tilde B}'
+ \frac{1}{\phi'} {\tilde C}' - \frac{2 {\cal H}}{\phi'} {\tilde A} + \nonumber\\
& & \left[ \frac{a^2 V_{,\phi}}{{\cal H} M_p^2} + \frac{6}{\phi'} {\cal H}^2 - \frac{\phi'^3}{2 M_p^4 {\cal H}^2}
-  \frac{\Delta}{\phi'} + \frac{2 \phi'}{M_p^2}  \right]  {\tilde B} + \left[ \frac{\cal H}{\phi'}+\frac{\phi'}{2 M_p^2 {\cal H}} \right]
{\tilde C} \Bigg\} \,,\nonumber\\  
& & \equiv J_v \,,
\end{eqnarray} 
where the l.h.s. is the standard term with 
\begin{equation} 
v  \equiv  a \delta \phi \;\;,\;\; z \equiv \frac{a \phi'}{\cal H} \nonumber\\
\end{equation} 
and where the sources are
\begin{eqnarray}\label{defss}
{\tilde A} & \equiv & \frac{a^2}{2} \left( T_\sigma \right)^i_i - \frac{3}{2} a^2 \Delta^{-1} \partial_i \partial_j \left( T_\sigma \right)^i_j\,, \nonumber\\ 
{\tilde B} & \equiv & a^2 \Delta^{-1} \partial_i \left( T_\sigma \right)^0_i \,,\nonumber\\ 
{\tilde C} & \equiv & a^2 \left( T_\sigma \right)^0_0 \,.
\end{eqnarray} 
The energy-momentum tensor of $\sigma$ receive contributions from its kinetic term and from its coupling to gravity
\begin{equation}
\left(T_{\sigma}\right)_{\mu\nu}=T_{\mu\nu}^{(M)}+T_{\mu\nu}^{(NM)},
\end{equation}
where
\begin{eqnarray}\label{minm}
& & T_{\mu\nu}^{(M)}\equiv    g_{\mu\nu}P-2P_{,X}\partial_{\mu}\sigma\partial_{\nu}\sigma\,,\quad\quad P\equiv\Lambda^{4}\left(\frac{-X}{M^{4}}\right)^{n}               \\\label{nminm}
& & T_{\mu\nu}^{(NM)}\equiv -\frac{\Lambda^{2}}{M^{4}} \Biggl\{ R(\sigmamnd) - 4 R_{(\mu\gamma}\nabla_{\nu)}\sigma\nabla^\gamma\sigma + \ggmn(\nabla\sigma)^2- 2\nabla_\gamma\nabla_\mu\sigma\nabla^\gamma\nabla_\nu\sigma + 2\nabla_\mu\nabla_\nu\sigma\square\sigma\nonumber\\
&&\quad\quad\quad\quad + \gmn\Bigr[ \nabla^\lambda\nabla^\delta\sigma\nabla_\lambda\nabla_\delta\sigma- (\square\sigma)^2 + 2\nabla^\alpha\sigma R_{\alpha\beta}\nabla^\beta\sigma\Bigr]\Biggl\}
\end{eqnarray}
Inserting (\ref{minm}) and (\ref{nminm}) into Eqs.~(\ref{defss}) and retaining the leading order terms in slow-roll and for $n\gg 1$, one arrives at the following result for the source
\begin{equation}
\frac{J_{v}}{a}\simeq -\frac{12}{\sqrt{2\epsilon}M_{P}}\frac{\Lambda^{2}H^{2}}{M^{4}}\partial_{i}\delta\sigma\partial_{i} \left[\delta\sigma-\frac{2}{\mathcal{H}}\delta\sigma^{'}\right]+\frac{J_{v}^{(NM)}}{a}\,,
\end{equation}
where 
\begin{eqnarray}
\frac{J_{v}^{(NM)}}{a}&\simeq &-\frac{1}{\sqrt{2\epsilon}M_P}
\frac{\Lambda^{2}H^{2}}{M^{4}}\frac{1}{\mathcal{H}^{2}}\Biggl\{ 6\Delta^{-1}\Bigl[( \Delta \delta\sigma')^2 - \partial_{i}\partial_{j}\dsg' \partial_{i}\partial_{j}\dsg'+  \Delta \delta\sigma \Delta \delta\sigma'' - \partial_{i}\partial_{j}\dsg \partial_{i}\partial_{j}\delta\sigma'' \Bigl]  \nonumber\\&-&2 \dij\Bigl[\partial_{i}\dsg' \partial_{j}\dsg'+\partial_{i}\partial_{j}\dsg \Delta\dsg-\partial_{i}\partial_{k}\dsg\partial_{j}\partial_{k}\dsg -\dsg'' \partial_{i}\partial_{j}\dsg \Bigl]-4\Bigl[\delta\sigma''\Delta\dsg+2 \delta\sigma'\Delta\dsg'\Bigl]\Biggl\} \nonumber\\
\end{eqnarray}
originates from the non-minimal coupling and $\dij\equiv \delta_{ij}-3 \Delta^{-1}\partial_{i}\partial_{j}$.

\section{Tadpole contribution to the tensor power spectrum}
\label{C}

In this section, we comment on the amplitude of the tadpole diagram for the tensor modes that arises from fourth order interactions of the kind $\gamma^{2}\delta\sigma^{2}$. \\
One may compute the interaction Hamiltonian to fourth order for interactions between two tensor modes and two spectator field fluctuations and use the in-in formalism to compute these diagrams. As an alternative, one could derive the non-linear equation of motion for the tensor modes and employ the Green's function method outlined in the main text. \\
The Hamiltonian to fourth order can be computed by expanding the Lagrangian in $\gamma^{2}\delta\sigma^{2}$ and, in addition to that, by accounting for the contributions that arise from the full third order Lagrangian \cite{Huang:2006eha}. The expansions are straightforward but we will not go through all the details and the intermediate results: the main purpose of this Appendix is to show that the tadpole does not provide a large enough contribution to the tensor power spectrum so as to affect the values of the tensor-to-scalar ratio that we derived from the two-vertex diagram and, more in general, as to affect the final results presented in Sec.~(\ref{sec:pheno}).\\

First of all, as explained in Sec.~(\ref{sec:zeta}), the higher the number of spatial derivatives acting on each spectator field fluctuations for each $\gamma^{2}\delta\sigma^{2}$ interaction, the bigger the $c_{s}$ enhancement one should expect. Given the form of the Lagrangian for $\sigma$, the maximum number of spatial derivatives that one can expect on each $\delta\sigma$ is one, so in this sense the interactions that maximize the $c_{s}$ enhancement are of the type $\mathcal{O}_{ij}[\gamma^{2}]\partial_{i}\delta\sigma\partial_{j}\delta\sigma$, with $\mathcal{O}_{ij}$ a generic operator acting on two tensor modes. One can, for instance, derive a contribution of this particular type from the spectator field quartic Lagrangian (after expanding and summing up the non-minimal and the minimal terms)
\ba\label{llll}
\mathcal{H}^{(4)}_{}\supset-\mathcal{L}^{(4)}_{}\supset a^{4}\left(\frac{\epsilon\Lambda^{2}H^{2}}{M^{4}}\right)\gamma_{ik}\gamma_{kj}\frac{\partial_{i}\delta\sigma\partial_{j}\delta\sigma}{a^{2}}.
\ea
By looking at the full Lagrangian and at the background solution for the spectator field, Eq.~(\ref{anz}), a dimensionless coupling constant $\Lambda^{2}H^{2}/M^{4}$ is to be expected for all the terms in the interaction Hamiltonian. An interaction as in Eq.~(\ref{llll})  can then be viewed as a typical contribution for the interactions in $\mathcal{H}^{(4)}[\gamma^{2}\delta\sigma^{2}]$. Notice that, as shown in the last part of this Appendix, the result for the tadpole diagram with an interaction that contains one spatial derivative acting on each $\delta\sigma$, leaves the final results of Section~(\ref{sec:pheno}) unaltered; this is true whether or not one takes a coupling constant with the $\epsilon$ suppression, i.e. $\epsilon\Lambda^{2}H^{2}/M^{4}$ or $\Lambda^{2}H^{2}/M^{4}$. \\

Finally, notice that estimating the amplitude of a tadpole diagram once the interactions have been identified (specifically once one knows what the coupling constants and the number of spatial derivatives acting on the spectator field fluctuations are), is a straightforward matter because, unlike the case of two-vertex loop diagrams, the $c_{s}$ dependence can be completely factored out of the integration; this is because the mode-functions do not have any dependence from the angle between the external and the internal momenta (a property of one-vertex diagrams).\\

We have all the ingredients necessary to compute the typical amplitude of the largest tadpole contributions to the tensor power spectrum
\ba\label{n2}
\langle \gamma\gamma \rangle_{tadpole}  \supset \frac{H^{4}}{M_{P}^{4}c_{s}^{3}k^{3}}.
\ea
It is clear from a comparison with the result for the two-vertex diagrams, Eq.~(\ref{tps}), that adding tadpole contributions would not affect the final results for the tensor power spectra that are presented in the main text.

\end{document}